\pdfoutput=1
\documentclass[12pt,a4paper]{article}

\usepackage{ifthen} 
\newboolean{pdflatex}
\setboolean{pdflatex}{true} 
 
\newboolean{articletitles}
\setboolean{articletitles}{true} 
 
\newboolean{uprightparticles}
\setboolean{uprightparticles}{false} 
 
\newboolean{inbibliography}
\setboolean{inbibliography}{false} 

\usepackage{microtype}
\usepackage{lineno}  
\usepackage{xspace} 
 
\usepackage{graphicx}  
\usepackage{color}
\usepackage{colortbl}
 
\usepackage{amsmath} 
\usepackage{amssymb}
\usepackage{amsfonts}
\usepackage{upgreek} 
 
\newcommand*\patchAmsMathEnvironmentForLineno[1]{%
\expandafter\let\csname old#1\expandafter\endcsname\csname #1\endcsname
\expandafter\let\csname oldend#1\expandafter\endcsname\csname
end#1\endcsname
 \renewenvironment{#1}%
   {\linenomath\csname old#1\endcsname}%
   {\csname oldend#1\endcsname\endlinenomath}%
}
\newcommand*\patchBothAmsMathEnvironmentsForLineno[1]{%
  \patchAmsMathEnvironmentForLineno{#1}%
  \patchAmsMathEnvironmentForLineno{#1*}%
}
\AtBeginDocument{%
\patchBothAmsMathEnvironmentsForLineno{equation}%
\patchBothAmsMathEnvironmentsForLineno{align}%
\patchBothAmsMathEnvironmentsForLineno{flalign}%
\patchBothAmsMathEnvironmentsForLineno{alignat}%
\patchBothAmsMathEnvironmentsForLineno{gather}%
\patchBothAmsMathEnvironmentsForLineno{multline}%
}
 
\usepackage{hyperref}    
\usepackage[all]{hypcap} 
 
\def\ux85 {UX85\xspace}

\ifthenelse{\boolean{uprightparticles}}%
{

 \def\Ppi         {\ensuremath{\uppi}\xspace}

 \def\PDelta      {\ensuremath{\Delta}\xspace}                 
 \def\PXi      {\ensuremath{\Xi}\xspace}                 
 \def\PLambda      {\ensuremath{\Lambda}\xspace}                 
 \def\PSigma      {\ensuremath{\Sigma}\xspace}                 
 \def\POmega      {\ensuremath{\Omega}\xspace}                 
 \def\PUpsilon      {\ensuremath{\Upsilon}\xspace}                 
 
 \def\PB      {\ensuremath{\mathrm{B}}\xspace}                 
                  
 \def\PD      {\ensuremath{\mathrm{D}}\xspace}

 \def\PK      {\ensuremath{\mathrm{K}}\xspace}

 \def\Pb      {\ensuremath{\mathrm{b}}\xspace}                 
 \def\Pc      {\ensuremath{\mathrm{c}}\xspace}

 \def\Pi      {\ensuremath{\mathrm{i}}\xspace}

}
{

 \def\Ppi         {\ensuremath{\pi}\xspace}

 \mathchardef\PDelta="7101
 \mathchardef\PXi="7104
 \mathchardef\PLambda="7103
 \mathchardef\PSigma="7106
 \mathchardef\POmega="710A
 \mathchardef\PUpsilon="7107
                  
 \def\PB      {\ensuremath{B}\xspace}                 
                  
 \def\PD      {\ensuremath{D}\xspace}

 \def\PK      {\ensuremath{K}\xspace}

 \def\Pb      {\ensuremath{b}\xspace}                 
 \def\Pc      {\ensuremath{c}\xspace}

 \def\Pi      {\ensuremath{i}\xspace}

}

\def\cquark    {\ensuremath{\Pc}\xspace}

\def\bquark    {\ensuremath{\Pb}\xspace}

\def\pion  {\ensuremath{\Ppi}\xspace}

\def\pipm  {\ensuremath{\pion^\pm}\xspace}

\def\kaon  {\ensuremath{\PK}\xspace}
  \def\Kbar  {\kern 0.2em\overline{\kern -0.2em \PK}{}\xspace}

\def\Kz    {\ensuremath{\kaon^0}\xspace}
\def\Kzb   {\ensuremath{\Kbar^0}\xspace}
\def\KzKzb {\ensuremath{\Kz \kern -0.16em \Kzb}\xspace}
\def\Kp    {\ensuremath{\kaon^+}\xspace}
\def\Km    {\ensuremath{\kaon^-}\xspace}

\def\KpKm  {\ensuremath{\Kp \kern -0.16em \Km}\xspace}

  \def\Dbar    {\kern 0.2em\overline{\kern -0.2em \PD}{}\xspace}
\def\D       {\ensuremath{\PD}\xspace}

\def\Dz      {\ensuremath{\D^0}\xspace}
\def\Dzb     {\ensuremath{\Dbar^0}\xspace}
\def\DzDzb   {\ensuremath{\Dz {\kern -0.16em \Dzb}}\xspace}
\def\Dp      {\ensuremath{\D^+}\xspace}
\def\Dm      {\ensuremath{\D^-}\xspace}

\def\DpDm    {\ensuremath{\Dp {\kern -0.16em \Dm}}\xspace}

\def\B       {\ensuremath{\PB}\xspace}
  \def\Bbar    {\kern 0.18em\overline{\kern -0.18em \PB}{}\xspace}
\def\Bb      {\ensuremath{\Bbar}\xspace}

\def\Bzb     {\ensuremath{\Bbar^0}\xspace}

\def\Bub     {\ensuremath{\B^-}\xspace}

\def\Bm      {\ensuremath{\Bub}\xspace}

  \def\Y#1S{\ensuremath{\PUpsilon{(#1S)}}\xspace}

\def\L {\ensuremath{\PLambda}\xspace}

\def\Lz {\ensuremath{\PLambda}\xspace}
\def\Lb      {\ensuremath{\Lz^0_\bquark}\xspace}

\def\Lc      {\ensuremath{\L_\cquark}\xspace}

\def\to                 {\ensuremath{\rightarrow}\xspace}

\def\AT#1     {\ensuremath{A_{\mathrm{T}}^{#1}}\xspace}           

\def\C#1      {\ensuremath{\mathcal{C}_{#1}}\xspace}                       
\def\Cp#1     {\ensuremath{\mathcal{C}_{#1}^{'}}\xspace}                    
\def\Ceff#1   {\ensuremath{\mathcal{C}_{#1}^{\mathrm{(eff)}}}\xspace}        
\def\Cpeff#1  {\ensuremath{\mathcal{C}_{#1}^{'\mathrm{(eff)}}}\xspace}       
\def\Ope#1    {\ensuremath{\mathcal{O}_{#1}}\xspace}                       
\def\Opep#1   {\ensuremath{\mathcal{O}_{#1}^{'}}\xspace}                    

\newcommand{\tev}{\ensuremath{\mathrm{\,Te\kern -0.1em V}}\xspace}
\newcommand{\gev}{\ensuremath{\mathrm{\,Ge\kern -0.1em V}}\xspace}
\newcommand{\mev}{\ensuremath{\mathrm{\,Me\kern -0.1em V}}\xspace}
\newcommand{\kev}{\ensuremath{\mathrm{\,ke\kern -0.1em V}}\xspace}
\newcommand{\ev}{\ensuremath{\mathrm{\,e\kern -0.1em V}}\xspace}
\newcommand{\gevc}{\ensuremath{{\mathrm{\,Ge\kern -0.1em V\!/}c}}\xspace}
\newcommand{\mevc}{\ensuremath{{\mathrm{\,Me\kern -0.1em V\!/}c}}\xspace}
\newcommand{\gevcc}{\ensuremath{{\mathrm{\,Ge\kern -0.1em V\!/}c^2}}\xspace}
\newcommand{\gevgevcccc}{\ensuremath{{\mathrm{\,Ge\kern -0.1em V^2\!/}c^4}}\xspace}
\newcommand{\mevcc}{\ensuremath{{\mathrm{\,Me\kern -0.1em V\!/}c^2}}\xspace}

\def\gsim{{~\raise.15em\hbox{$>$}\kern-.85em
          \lower.35em\hbox{$\sim$}~}\xspace}
\def\lsim{{~\raise.15em\hbox{$<$}\kern-.85em
          \lower.35em\hbox{$\sim$}~}\xspace}

\def\tell1  {TELL1\xspace}
\def\ukl1   {UKL1\xspace}

\def\Sb      {\ensuremath{\PSigma_\bquark}\xspace}

\usepackage{cite} 
\usepackage{mciteplus}
\begin{document}

\begin{titlepage}

\vspace*{-1.5cm}
\vspace*{1.5cm}
\hspace*{-5mm}\begin{tabular*}{16cm}{lc@{\extracolsep{\fill}}r}
 & & \today \\ 
\end{tabular*}

\vspace*{4.0cm}

{\bf\boldmath\huge
\begin{center}
Method of Studying \Lb decays with one missing particle
\end{center}
}

\vspace*{2.0cm}

\begin{center}
Sheldon Stone and Liming Zhang 
\bigskip\\
{\it\footnotesize
Physics Department
Syracuse University, Syracuse, NY, USA 13244-1130\\

}
\end{center}

\vspace{\fill}

\begin{abstract}
 \noindent
A new technique is discussed that can be applied to \Lb baryon decays where decays with one missing particle can be discerned from background and their branching fractions determined, along with other properties of the decays. Applications include measurements of the CKM elements $|V_{ub}|$ and $|V_{cb}|$,  and detection of any exotic objects coupling to $b\to s$ decays, such as the inflaton. Potential use of $\overline{B}^{0**}\to\pi^+B^-$ and $\overline{B}_s^{0**}\to K^+B^-$ to investigate \Bm decays is also commented upon.
\end{abstract}

\vspace*{2.0cm}

\begin{center}
  Submitted to Advances in High Energy Physics 
\end{center}

\vspace{\fill}

\end{titlepage}

\pagestyle{empty}  



\setcounter{page}{2}
\mbox{~}


\pagestyle{plain} 
\setcounter{page}{1}
\pagenumbering{arabic}


\section{Introduction}

Detection of $b$-flavored hadron decays with one missing neutral particle, such as a neutrino, is important for many measurements and searches. These include semileptonic decays, such as $\Bb\to D\mu^-\overline{\nu}$, $\Bb\to \pi\mu^-\overline{\nu}$,  and any exotic long lived particles that could be produced in decays such as $\Bb\to X \chi$, where the $X$ is any combination of detected particles and the $\chi$ escapes the detector.\footnote{In this paper mention of a particular decay mode implies the use of the charge-conjugated mode as well.}   These measurements are possible at an $e^+e^-$ collider operating at the $\Upsilon(4S)$. Since $\Upsilon(4S)\to B\overline{B}$, fully reconstructing either the \B or the \Bb determines the negative of the initial four-momentum of the other. With this information it is possible to measure final states where one particle is not detected, such as a neutrino. To implement this procedure, taking the \Bb  to be fully reconstructed,  the missing mass-squared, $m_x^2$, is calculated including the information on the initial $B$ four-momentum  and measurements  of the found $X$ particles as  
\begin{equation}
\label{eq:mxsq}
m_x^2=(E_B-E_X)^2-(\overrightarrow{p}\!\!_{B}-\overrightarrow{p}\!\!_X)^2, 
\end{equation}
where $E$ and $\overrightarrow{p}$ indicate energy and three-momentum, respectively. Peaks in $m_x^2$ would be indicative of single missing particles in the $B$ decay. 

A related example is charm semileptonic decays with a missing neutrino. Determinations of branching fractions and form-factors  have been carried out in fixed target experiments, exploiting the measured direction of the charmed hadron and assuming that the missing particle has zero mass, which leads to a two-fold ambiguity in the neutrino momentum calculation \cite{Aitala:1998ey}. 
 If the charm decay particle is a $D^0$, extra constraints can be imposed on its decay requiring it to be produced from a $D^{*+}$ in the decay $D^{*+}\to \pi^+ D^0$. This leads to more constraints than unknowns, and is quite useful for rejecting backgrounds  \cite{Link:2004uk,*Agostino:2004na}.

Interesting decays of the $\Lb$ baryon also exist, but investigations are not feasible in the $\Upsilon(4S)$ energy region.  
Potential studies include determination of the CKM matrix element $|V_{cb}|$, possible using $\Lb\to\Lc^+ \ell^-\overline{\nu}$ decays and $|V_{ub}|$ using the $\Lb\to p \ell^-\overline{\nu}$ mode. 

\begin{figure}[b]
\begin{center}
\includegraphics[width=6in]{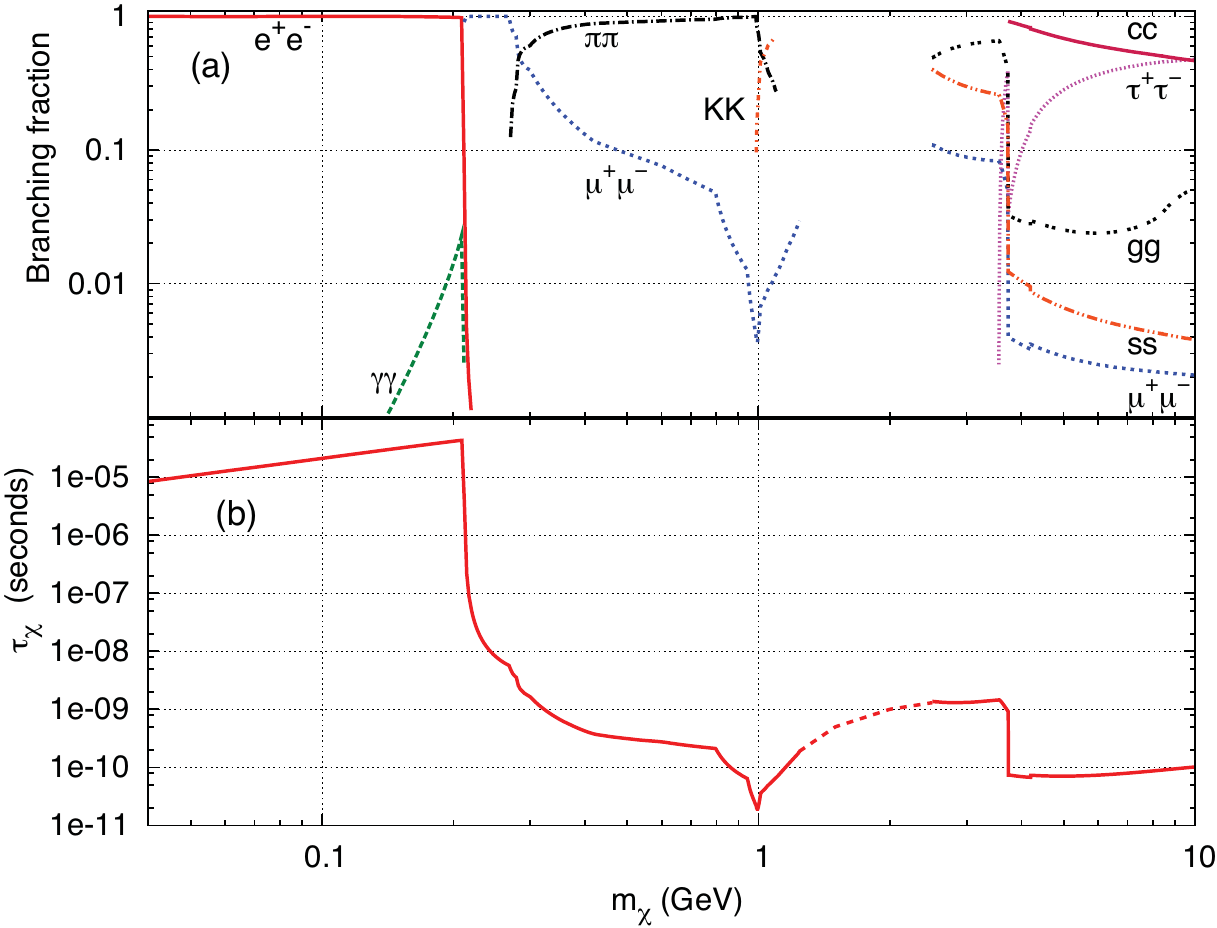}
\end{center}\label{branch-life}
\vskip -0.5cm
\caption{Predictions from Ref.~\cite{Bezrukov:2009yw}. (a)
    Inflaton branching ratios to various two-body final states 
    as functions of inflaton mass for
    $m_\chi<1.5$\,GeV. Above 2.5\,GeV only quark-antiquark and dilepton modes are predicted. In the intermediate region
   no reliable prediction is given. 
   (b)
     Inflaton lifetime $\tau_\chi$ as a function of the inflaton
    mass $m_\chi$.  The
    lifetime can be up to two times smaller, depending on model-dependent parameters.
  }
\end{figure}

Neutral particles that have not yet been seen could be searched for, even if they are  stable or have long enough lifetimes that they would have only a very small fraction of their decays inside the detection apparatus.  One example of such a possibly long-lived particle is the ``inflaton." This particle couples to a scalar field and is responsible for cosmological inflaton. Bezrukov and Gorbunov predicted branching fractions and decay modes of inflatons, $\chi$, in $\Bb$ meson decays \cite{Bezrukov:2009yw} using a specific model,  which is a particular version of the simple chaotic inflation with a quartic potential and having the inflaton field coupled to the SM Higgs boson via a renormalizable operator.  For $\Bb\to \chi X_s$ decays the branching fraction is 
\begin{align}
  \label{BrhX}
  {\cal{B}}( \Bb\to \chi X_s)
    & \simeq 0.3 \frac{\left|V_{ts}V_{tb}^* \right|^2}{\left| V_{cb}\right|^2}  \left( \frac{m_t}{M_W}
      \right)^4 \left( 1-\frac{m_\chi^2}{m_b^2}\right)^2 \theta^2   \\\nonumber
    & \simeq 10^{-6} \cdot\left(
      1-\frac{m_\chi^2}{m_b^2}\right)^2 \left( \frac{\beta}{\beta_0}\right) \left(
      \frac{\rm 300~MeV}{m_\chi}\right)^2,
\end{align}
where $X_s$ stands for strange meson channels mostly saturated by a sum
of $K$ and $K^*(890)$ mesons, $m_{\chi}$ and $m_t$, the inflaton and top quark masses, respectively. The model parameters are $\theta$, $\beta$ and $\beta_0$, where 
$\beta/\beta_0\approx{\cal{O}}(1).$ Their inflaton branching fraction predictions are shown in Fig.~\ref{branch-life}(a). 
The branching fractions are quite similar for $\Lb$ decays. The $\Lb\to pK^-\chi$ channel would seem to be the most favorable, since the $\Lb$ decay point could be accurately determined from the $p K^-$  
vertex. The mass dependent inflaton branching fraction predictions for different decay modes are shown in Fig.~\ref{branch-life}(a). Collider searches that rely on directly detecting the inflaton decay products may not be sensitive to lifetimes much above a few times 1~ns~\cite{Aaij:2014aba}, because the particles mostly decay outside of the detector, while searches that could be done inclusively, e.g., without detecting the inflaton decay products, would be independent of this restriction.

Use of $\Lb$ decays in measuring CKM matrix elements as well as new particle searches has been not as fruitful as in $B$ meson decays because $e^+e^-$ machines have access only to the lighter $B$ mesons. In addition, absolute branching fraction determinations have been made difficult by the relatively large uncertainty on ${\cal{B}}(\Lc^+\to p K^- \pi^+)$. Recently, the Belle collaboration reduced this uncertainty from 25\% to about 5\%, allowing for measurements with much better precision~\cite{Zupanc:2013iki}.

Inclusive decay searches using $\Lb$ baryons can be made at high energy colliders if it were possible to find a way to estimate the $\Lb$ momentum. The $\Lb$ direction is measured by using  its finite decay distance. To get an estimate of the $\Lb$ energy we can use $\Lb$'s that come from $\PSigma_b^{\pm}\to \pi^{\pm}\Lb$ and $\PSigma_b^{*\pm}\to \pi^{\pm}\Lb$ decays. The $\PSigma_b^{(*)\pm}$ states were found by the CDF collaboration \cite{CDF:2011ac}. Their masses and widths are consistent with theoretical predictions \cite{Karliner:2008sv}.

The $\Lb$ energy is determined from the measurement of  the $\pi^{\pm}$ from the $\PSigma_b^{(*)}$ decay along with the $\Lb$ direction. Let us assume we have a pion from the $\PSigma_b^{(*)}$ decay. Then
\begin{equation}
\label{eq:mdiff}
m_{\Sb^{(*)}}^2=(E_\pi + E_{\Lb})^2-(\overrightarrow{p}\!\!_\pi+\overrightarrow{p}\!\!_{\Lb})^2,
\end{equation}
and after some algebraic manipulations we find
\begin{eqnarray}
\label{eq:plb}
|p_{\Lb}|&=&(-b\pm\sqrt{b^2-4ac})/(2a) \\\nonumber
a&=&4(E^2_\pi-p^2_\pi\cos^2\theta) \\ \nonumber
b&=&-4p_\pi\Delta^2_m\cos\theta\\\nonumber
c&=&4E^2_\pi m^2_{\Lb}- \Delta^4_m \\\nonumber
\Delta^2_m&=&m^2_{\Sb^{(*)}}-m^2_\pi-m^2_{\Lb} , 
\end{eqnarray}
where $\cos\theta$ is the measured angle between the pion and the $\Lb$, and $m_{\Sb^{(*)}}$ indicates either the $\PSigma_b$ or $\PSigma_b^*$ mass. With the measured $\Lb$ direction and $\Lb$ energy Eq.\,(\ref{eq:mxsq}) can now be used to find decays with any number of detected and one missing particle. Two possible solutions result because of the $\pm$ sign in the first line of Eq.~\ref{eq:plb}. In similar studies one solution is often unphysical. This was seen, for example, using $\D^{*+}\to\pi^+ D^0$, $D^0\to K^-\pi^+\pi^+\pi^-$ decays with one missing pion \cite{Aaij:2012cy}. The resolution in  $m_x^2$ depends on several quantities including the measurement uncertainties on momentum of the final state particles and the $\Lb$ direction, so it may be advantageous for analyses to select long-lived decays at the expense of statistics. The relatively long $\Lb$ lifetime of about 1.5~ps is helpful in this respect \cite{Aaij:2013oha,*Aaij:2014owa}.

\begin{figure} [!b]
\begin{center} 
\vspace{-4cm}
  \includegraphics[width=0.99\textwidth]
  {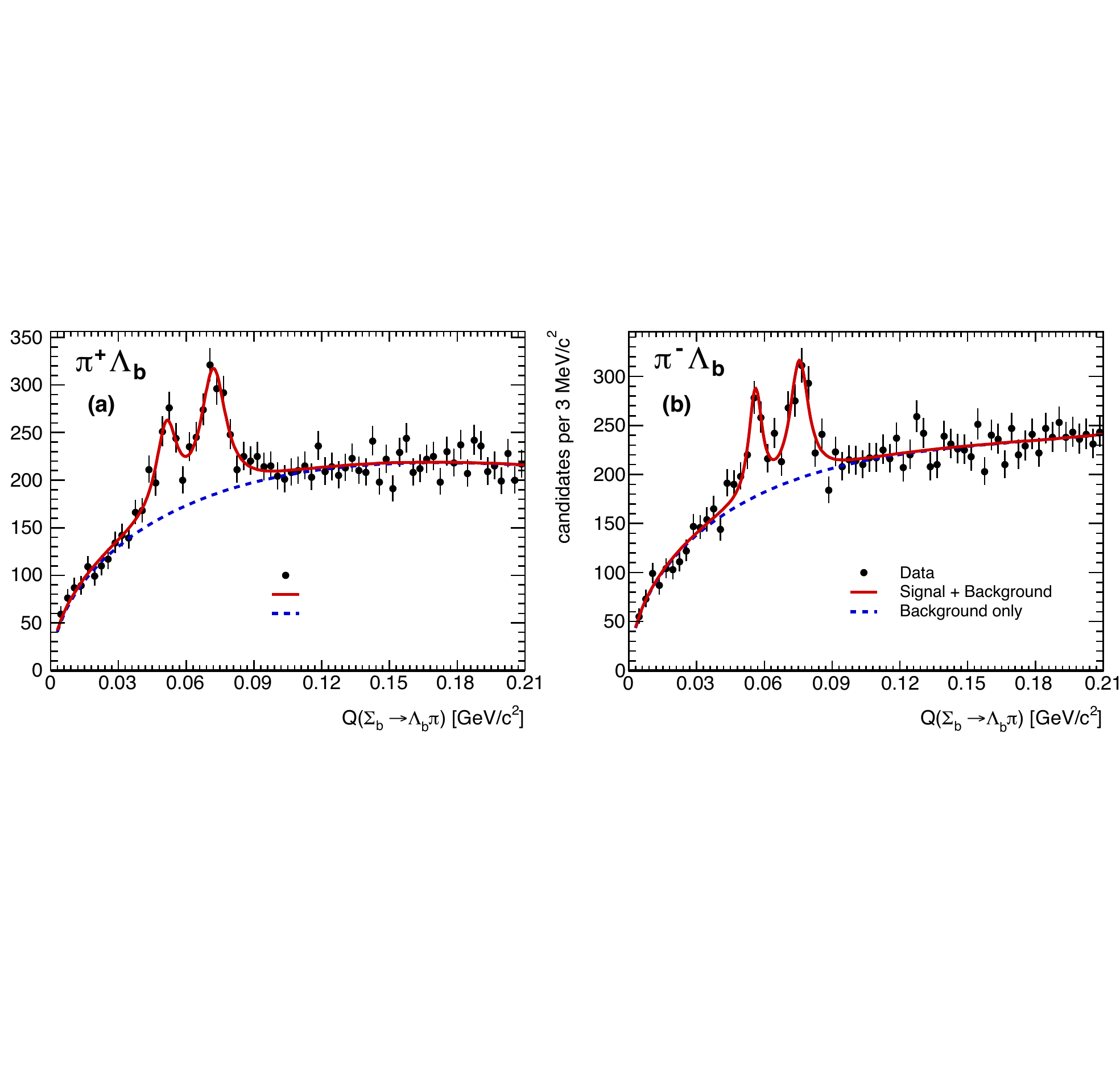}  
\vspace*{-4.5cm}
  \caption{ The \({Q= M(\Lb\pipm)-M(\Lb)-m_{\pi}}\) spectrum for 
             candidates with the projection of the
            corresponding unbinned likelihood fit superimposed, (a) for  $\pi^+\Lb$ and (b) for $\pi^-\Lb$ candidates. (From Ref.~\cite{CDF:2011ac}).}
\label{fig:signal-sgbmp}
\end{center}
\end{figure}

The $\Sb^{(*)\pm}$ states have only been seen by CDF \cite{CDF:2011ac}. Their data are shown in Fig.~\ref{fig:signal-sgbmp}, and listed in Table~\ref{tab:fitresults}.

\begin{table}[h]
\begin{center}
\caption{ Summary of the results of the fits to the 
          \({Q = M(\Lb\pipm)-M(\Lb)-m_{\pi}}\) spectra from CDF~ \cite{CDF:2011ac}.  
                    \label{tab:fitresults} }
  \begin{tabular}{lccc}
\hline
\hline 
    State        & \(Q\) value,   & Natural width,      & Yield \\
                 & MeV        & \(\Gamma_{0}\), MeV  &  \\
\hline
   ${\Sb^-}$\rule{0pt}{2.4ex}     & {\({56.2}_{-0.5}^{+0.6} \)} & {\({4.9}_{-2.1}^{+3.1} \)} & {\(340_{-70}^{+90}  \)} \\
    ${\Sb^{*-}}$ & {\({75.8}\pm{0.6} \)}       & {\({7.5}_{-1.8}^{+2.2} \)} & {\(540_{-80}^{+90}  \)} \\
    ${\Sb^+}$   & {\({52.1}_{-0.8}^{+0.9}\)}  & {\({9.7}_{-2.8}^{+3.8} \)} & {\(470_{-90}^{+110} \)} \\
    ${\Sb^{*+}}$ & {\({72.8}\pm{0.7}\)}        & {\({11.5}_{-2.2}^{+2.7}\)} & {\(800_{-100}^{+110}\)} \\
\hline
\hline
\end{tabular}
\end{center}
\end{table}

\section{Potential measurements}

Although there is no measurement of the relative $\Sb^{(*)\pm}/\Lb$ production cross-section, $r_{\PSigma\Lambda}$, one might imagine that the production ratio would be close to unity. The pions from the  $\Sb^{(*)\pm}$ decays have relatively low momenta, so their detection efficiencies could be small. Although CDF does not report a value for the production ratio, the number of seen signal events gives an observed value of $r_{\PSigma\Lambda}$ equal to 13\%. This is certainly a useful sample.  Backgrounds will be an issue, however, as the CDF data do show a substantial amount of non-resonant combinations under the signal peaks, but this will not prevent searches, just limit their sensitivities with a given data sample. 

Measurement of $|V_{cb}|$, determined using $\Lb\to\Lc^+ \ell^-\overline{\nu}$  decays with $\Lc^+\to pK^-\pi^+$ would provide an important cross-check on this important fundamental parameter, especially when updated lattice gauge calculations become available \cite{Bowler:1997ej}.  This measurement is not subject to the uncertainty on  ${\cal{B}}(\Lc^+\to pK^-\pi^+)$ provided that the total number of $\Lb$ events in the event sample is determined using the same branching fraction \cite{Aaij:2014jyk}. The LHCb determination of the ratio of $\Lb$ to \Bzb production, for example, uses the $\Lc^+\to pK^-\pi^+$ decay mode \cite{LHCb:2013lka}, and then the absolute number of $\Lb$ events produced is found by measuring the \Bzb rate in a channel with a known branching fraction. The branching ratio for the channel $\Lb\to\Lc^+ \ell^-\overline{\nu}$ can be determined using Eq.\,(\ref{eq:mxsq}) using the measured value for the $\Lb$ energy determined by using Eq.\,(\ref{eq:plb}); a signal would appear near $m_x^2$ equal to zero. To determine the four-momentum transfer squared from the $\Lb$ to the $\Lc^+$ a similar procedure as used in the decay sequence $\D^{*+}\to D^0\pi^+$, $D^0\to K^{*-}\ell^+\nu$ can be implemented \cite{Link:2004uk,*Agostino:2004na}.  
In this procedure, the neutrino mass is set to zero,
\begin{equation}\label{eq:mxz}
(E_{\Lb}-E_{X})^2-(\vec{p}_{\Lb}-\vec{p}_{X})^2=m_x^2=0,
\end{equation}
where $X$ represents the sum of  $\Lc^+$ and $\ell^-$  energies and momenta.  Eq.\,(\ref{eq:mdiff}) and Eq.\,(\ref{eq:mxz}) can be used as two constraint equations with one unknown variable $|p_{\Lb}|$. 

Measurement of $|V_{ub}|$ using $\Lb\to p  \ell^-\overline{\nu}$ decays is subject to the uncertainty on ${\cal{B}}(\Lc^+\to pK^-\pi^+)$ , but here the current precision of 5\% on this branching fraction is sufficient. Theoretical calculations of the decay width from the lattice gauge calculations done in a limited four-momentum transfer range \cite{Detmold:2013nia}, light cone sum rules \cite{Khodjamirian:2011jp,Wang:2009hra,*Azizi:2009wn,*Huang:2004vf},  and QCD sum rules \cite{MarquesdeCarvalho:1999ia,*Huang:1998rq}
can be used to extract $|V_{ub}|$. The $ p  \ell^-\overline{\nu}$ final state is subject to backgrounds from $N^*\ell^-\overline{\nu}$, where $N^*\to p \pi^0$, that are difficult to eliminate and thus the use of the $\Sb^{(*)\pm}\to \pi^{\pm}\Lb$ decay sequence may be crucial. The decay sequence constraint can also possibly help measure the branching fraction for $\Lb\to \Lc^{(*)+}\tau^-\overline{\nu}$ decays as measurements in the \Bb meson system of analogous decays are somewhat larger than Standard Model predictions \cite{Lees:2013uzd,*Adachi:2009qg,*Bozek:2010xy}. 

Particles characteristic of scalar fields such as inflatons or dilatons can be searched for in $\Lb$ decays. It is also possible to search for Majorana neutrinos through a process similar to that used for searches in $B^-\to\mu^-\mu^-\pi^+$ decays \cite{Aaij:2014aba,Aaij:2012zr,*Aaij:2011ex}, where the Majorana neutrino, $\nu_M$, decays into a $\mu^-\pi^+$ pair.  The initial quark content of the $\Lb$ is $bud$. The $b$-quark can annihilate with a $\overline{u}$-quark from a $u\overline{u}$ pair arising from the vacuum into a virtual $W^-$ leaving a $uud$ system that can form a $p$. The virtual $W^-$ then can decay into $\mu^-$ in association with a Majorana neutrino that can transform to its own anti-particle and decay into $\mu^-$ and a virtual $W^+$. In the analogous case to the $B^-\to\mu^-\mu^-\pi^+$ decay, the $W^+$ would decay into a $\pi^+$, however here we do not have to detect the Majorana decays, so we can look for the decay $\Lb\to p\mu^-\nu_M$ independently of the $\nu_M$ decay mode or lifetime. Other mechanisms for Majorana neutrino production discussed in Ref.~\cite{Castro:2013jsn} for $B^-$ decays when adopted to $\Lb$ decays, would also lead to the $p\mu^-\nu_M$ final state.

A more mundane search can be considered for $\Lb$ decays into non-charmed final states containing $\PSigma^{\pm}$ light baryons; these have been proposed  for flavor SU(3) tests \cite{Gronau:2013mza}. Since the largest decay modes are  $\PSigma^-\to n \pi^-$, and $\PSigma^+\to n\pi^+$ or $p\pi^0$, there is always a missing neutron in the $\PSigma^-$ decay, while the $\PSigma^+$, in principle, can be detected in the $p\pi^0$ mode. The method suggested here can be adopted to search for both $\PSigma^-$ and $\PSigma^+$ baryons in $\Lb$ decays.  

Note that similar methods can be applied to $B^-$ decays by the use of the $\overline{B}^{0**}\to \pi^+B^-$ decay sequence. The measured production ratio of $\left(\overline{B}^{0**}\to \pi^+ B^-\right)/B^-$ is about 15\%, but the $B^{0**}$'s have widths of about 130~MeV which introduces very large backgrounds that have thus far precluded their use. Another possible source of tagged $B^-$ events is the decays of $\overline{B}_s^{0**}$ mesons into a $K^+B^-$ that would have the advantage of a charged kaon tag, and have a much narrower width.

In conclusion, we propose a new method of analyzing $\Lb$ decays  into one missing particle, where the $\Lb$ is part of a detected $\Sb^{(*)\pm}\to \pi^{\pm}\Lb$ decay that provides additional kinematic constraints. This method may be useful for studies of CKM elements and searches for new particles such as inflatons, dilatons or Majorana neutrinos. Thus, investigations of $\Lb$ decays may present a unique opportunity in the study of $b$-flavored hadron decays.

\section*{Acknowledgements}
We are grateful for the support of the  National Science Foundation, and discussions with Jon Rosner, and many of our LHCb colleagues.

\ifx\mcitethebibliography\mciteundefinedmacro
\PackageError{LHCb.bst}{mciteplus.sty has not been loaded}
{This bibstyle requires the use of the mciteplus package.}\fi
\providecommand{\href}[2]{#2}

\end{document}